 \documentclass[final,5p,times,twocolumn]{elsarticle}

 \usepackage{graphics}
 \usepackage{graphicx}

\usepackage{amssymb}
\usepackage{amsmath}

\journal{Journal of Molecular Spectroscopy}

\begin{document}

\begin{frontmatter}



\title{Determination of accurate rest frequencies and hyperfine structure 
       parameters of cyanobutadiyne, HC$_5$N}


\author[Koeln]{Thomas~F. Giesen\fnref{PA2}} 
\fntext[PA2]{Present address: Laborastrophysik, Universit\"at Kassel, 34132 Kassel, Germany}
\author[Mainz,Karlsruhe]{Michael E. Harding\corref{cor}} 
\ead{michael.harding@kit.edu}
\author[Mainz]{J\"urgen Gauss} 
\author[Hann]{Jens-Uwe Grabow} 
\author[Koeln]{Holger S.P.~M\"uller\corref{cor}}
\ead{hspm@ph1.uni-koeln.de}
\cortext[cor]{Corresponding author.}

\address[Koeln]{I.~Physikalisches Institut, Universit{\"a}t zu K{\"o}ln, 
  Z{\"u}lpicher Str. 77, 50937 K{\"o}ln, Germany}
\address[Mainz]{Department Chemie, Johannes Gutenberg-Universit{\"a}t Mainz, 
Duesbergweg 10-14, 55128 Mainz, Germany}
\address[Karlsruhe]{Institut f\"ur Physikalische Chemie, Karlsruher Institut 
         f\"ur Technologie (KIT), Kaiserstra{\ss}e 15, 76131 Karlsruhe, Germany}
\address[Hann]{Institut f\"ur Physikalische Chemie und Elektrochemie, 
Lehrgebiet A, Leibniz-Universit\"at Hannover, Callinstr. 3-3A, 
30167 Hannover, Germany}

\begin{abstract}

Very accurate transition frequencies of HC$_5$N were determined between 5.3 and 21.4~GHz 
with a Fourier transform microwave spectrometer. The molecules were generated by passing 
a mixture of HC$_3$N and C$_2$H$_2$ highly diluted in neon through a discharge valve 
followed by supersonic expansion into the Fabry-Perot cavity of the spectrometer. 
The accuracies of the data permitted us to improve the experimental $^{14}$N nuclear 
quadrupole coupling parameter considerably and the first experimental determination of 
the $^{14}$N nuclear spin-rotation parameter. The transition frequencies are also well 
suited to determine in astronomical observations the local speed of rest velocities 
in molecular clouds with high fidelity. The same setup was used to study HC$_7$N, albeit 
with modest improvement of the experimental $^{14}$N nuclear quadrupole coupling parameter. 
Quantum chemical calculations were carried out to determine $^{14}$N nuclear quadrupole 
and spin-rotation coupling parameters of HC$_5$N, HC$_7$N, and related molecules. 
These calculations included evaluation of vibrational and relativistic corrections to the 
non-relativistic equilibrium quadrupole coupling parameters; their considerations improved 
the agreement between calculated and experimental values substantially. 

\end{abstract}


\end{frontmatter}




\section{Introduction}
\label{introduction}

Cyanopolyynes H(C$\equiv$C)$_n$CN occur abundantly in space, in particular the shorter members. 
Molecules up to cyanooctatetrayne, HC$_9$N ($n = 4$), were detected \cite{HC9N_det_1978}. 
The next longer member, HC$_{11}$N, has not yet been found in space \cite{HC11N_non-det_2016}. 
Isotopic species with D or with one $^{13}$C were detected up to HC$_7$N 
\cite{HC5_7N_isos_TMC-1_2018}, $^{15}$N isotopologs up to HC$_5$N \cite{HC5N-15_TMC-1_2017}, 
even all three isotopomers of HC$_3$N with two $^{13}$C were observed astronomically 
\cite{PN_K4-47_2019}. Measurements of the HC$_3$N species and the isotopomers with one 
$^{13}$C were frequently used to determine $^{12}$C/$^{13}$C ratios in various objects, 
such as the protoplanetary nebula CRL618 \cite{CRL618_12C-13C_etc_2003,CRL618_abundances_2007} 
or in starless cores, where differences in the $^{12}$C/$^{13}$C ratios were found 
\cite{HC3N_13C-fractionation_2017}.

Cyanobutadiyne, HC$_5$N, also known as cyanodiacetylene or pentadiynenitrile, was detected as 
early as 1976 toward the high-mass star-forming region Sagittarius B2 close to the Galactic 
center \cite{HC5N_det_1976}. It was found soon thereafter in the dark and dense core 
Heile's Cloud 2 \cite{HC5N_dark_cloud_1977}, nowadays better known as Taurus Molecular Cloud~1 
or short as TMC-1 \cite{name_TMC-1_1978}. Cyanohexatriyne, HC$_7$N, also known as 
cyanotriacetylene or heptatriynenitrile, was discovered in that source \cite{HC7N_det_1978}. 
Both molecules were also found early in the circumstellar envelope of the famous carbon 
rich asymptotic giant branch star CW~Leonis, also referred to as IRC+10216 
\cite{HC5N_HC7N_CW-Leo_1978}.

Molecules up to HC$_{17}$N ($n = 8$) were investigated by rotational spectroscopy 
\cite{HC15_17N_rot_1998}. Alexander et al. were the first to investigate the rotational spectrum 
of HC$_5$N \cite{HC5N_rot_dip_etc_1976}. They assigned ground state rotational spectra of eight 
isotopic species in the microwave (MW) region from which they determined structural parameters. 
They also determined the $^{14}$N nuclear quadrupole coupling parameter $eQq$(N) and the dipole 
moment of the main isotopic species. Winnewisser et al. improved $eQq$(N) \cite{HC5N_eQq_1978} 
and expanded the assignments into the millimeter wave (mmW) region \cite{HC5N_mmW_1982}. 
Bizzocchi et al. recorded the spectra of HC$_5$N and DC$_5$N in the mmW and sub-mmW regions 
\cite{HC5N_isos_v0_2004}. Assignments for these two isotopologs were extended to 460~GHz. 
They also analyzed spectra of singly substituted isotopic variants of both isotopologs 
containing one $^{13}$C or $^{15}$N, and derived from the rotational parameters a 
semi-experimental equilibrium structure.

Kirby et al. analyzed the ground state rotational spectrum of HC$_7$N in the MW region 
\cite{HC7N_rot_1980}. McCarthy et al. subjected the main isotopic species as well as all singly 
substituted ones to a Fourier transform (FT) MW spectroscopic study and determined $eQq$(N) 
for all of them \cite{HC7_9_11N_rot_eqQ_etc_2000}. Similar studies were carried out for 
HC$_9$N and HC$_{11}$N, and ground state effective structural parameters for all three 
molecules. Soon thereafter, Bizzocchi et al. extended the assignments of the main isotopic 
species in its ground and several low-lying vibrational states into the mmW region 
\cite{HC7N_mmW_2004}.

The aim of the present work is twofold. The first target was the improvement of the hyperfine 
structure (HFS) parameters of HC$_5$N and potentially of HC$_7$N, and secondly, we wanted 
to investigate how accurately HFS parameters can be evaluated by high-level quantum-chemical 
calculations, and in particular trends among related molecules. An earlier study of isotopic 
species associated with DC$_3$N showed that very good agreement can be achieved for the nuclear 
quadrupole parameters in high-accuracy coupled-cluster calculations by employing large basis sets 
together with vibrational corrections \cite{DC3N_rot_etc_2008}.


\section{Experimental details}
\label{exptl_details}

Spectra between 5 and 22~GHz were recorded at the Leibniz-Universit\"at in Hannover employing 
a supersonic-jet Fourier transform microwave (FTMW) spectrometer \cite{FTMW_1981} in the 
coaxially oriented beam-resonator arrangement (COBRA) \cite{COBRA_1990} which combines a very 
sensitive setup with an electric discharge nozzle \cite{harvard_setup_w-JUG_2005}. 
Cyanobutadiyne was generated by passing a mixture of equal amounts of 1\% HC$_3$N in neon 
and 1\% C$_2$H$_2$ in neon at a pressure of $\sim$100~kPa through the discharge nozzle and 
expanding the products into the cavity of the spectrometer. Cyanohexatriyne was obtained 
initially by the same procedure and subsequently by modifying the ratio of diluted HC$_3$N 
and C$_2$H$_2$ from 1~:~1 to 1~:~2. Uncertainties of frequencies of 0.3~kHz and below can be 
achieved under favorable conditions \cite{i-PrCN_rot_etc_2011,2-CAB_rot_2017}.


\begin{table}
\begin{center}
\caption{Rotational transitions of HC$_5$N with quantum numbers $J$ and $F$ recorded in the present study, 
         their frequencies (MHz), uncertainties Unc. (kHz), residuals O$-$C (kHz) between observed 
         transition frequencies and those calculated from the global fit.}
\label{HC5N-exp-data}
\begin{tabular}[t]{ccr@{}lr@{}lr@{}l}
\hline 
$J' - J''$ & $F' - F''$ &  \multicolumn{2}{c}{Frequency} &  \multicolumn{2}{c}{Unc.} &  \multicolumn{2}{c}{O$-$C} \\
\hline
 2 $-$ 1 & 2 $-$ 2 &  5324&.04642 & 0&.20 &    0&.05 \\
         & 1 $-$ 0 &  5324&.26054 & 0&.20 & $-$0&.06 \\
         & 2 $-$ 1 &  5325&.32957 & 0&.20 & $-$0&.24 \\
         & 3 $-$ 2 &  5325&.42182 & 0&.20 &    0&.03 \\
         & 1 $-$ 2 &  5326&.18468 & 0&.20 &    0&.20 \\
         & 1 $-$ 1 &  5327&.46808 & 0&.20 &    0&.16 \\
 3 $-$ 2 & 3 $-$ 3 &  7986&.61752 & 0&.20 &    0&.03 \\
         & 2 $-$ 1 &  7987&.77870 & 0&.20 &    0&.02 \\
         & 3 $-$ 2 &  7987&.99286 & 0&.20 & $-$0&.05 \\
         & 4 $-$ 3 &  7988&.04414 & 0&.20 &    0&.00 \\
         & 2 $-$ 3 &  7988&.54163 & 0&.20 &    0&.27 \\
         & 2 $-$ 2 &  7989&.91689 & 0&.20 &    0&.11 \\
 4 $-$ 3 & 4 $-$ 4 & 10649&.22745 & 0&.20 &    0&.27 \\
         & 3 $-$ 2 & 10650&.56191 & 0&.20 &    0&.06 \\
         & 4 $-$ 3 & 10650&.65418 & 0&.20 &    0&.34 \\
         & 5 $-$ 4 & 10650&.68658 & 0&.20 &    0&.04 \\
         & 3 $-$ 3 & 10652&.48571 & 0&.20 & $-$0&.02 \\
 5 $-$ 4 & 5 $-$ 5 & 13311&.85257 & 0&.20 &    0&.06 \\
         & 4 $-$ 3 & 13313&.26074 & 0&.20 &    0&.10 \\
         & 5 $-$ 4 & 13313&.31208 & 0&.20 &    0&.21 \\
         & 6 $-$ 5 & 13313&.33461 & 0&.20 &    0&.00 \\
         & 4 $-$ 4 & 13315&.09264 & 0&.20 &    0&.11 \\
 6 $-$ 5 & 6 $-$ 6 & 15974&.48392 & 0&.20 & $-$0&.29 \\
         & 5 $-$ 4 & 15975&.93351 & 0&.20 & $-$0&.08 \\
         & 6 $-$ 5 & 15975&.96655 & 0&.20 &    0&.25 \\
         & 7 $-$ 6 & 15975&.98302 & 0&.20 & $-$0&.03 \\
         & 5 $-$ 5 & 15977&.71410 & 0&.20 & $-$0&.15 \\
 7 $-$ 6 & 7 $-$ 7 & 18637&.11750 & 0&.20 & $-$0&.05 \\
         & 6 $-$ 5 & 18638&.59378 & 0&.20 &    0&.12 \\
         & 7 $-$ 6 & 18638&.61629 & 0&.20 & $-$0&.10 \\
         & 8 $-$ 7 & 18638&.62933 & 0&.20 &    0&.06 \\
         & 6 $-$ 6 & 18640&.34150 & 0&.20 & $-$0&.11 \\
 8 $-$ 7 & 8 $-$ 8 & 21299&.74993 & 0&.50 &    0&.23 \\
         & 7 $-$ 6 & 21301&.24497 & 0&.20 &    0&.30 \\
         & 8 $-$ 7 & 21301&.26134 & 0&.20 & $-$0&.08 \\
         & 9 $-$ 8 & 21301&.27125 & 0&.20 & $-$0&.40 \\
         & 7 $-$ 7 & 21302&.96959 & 0&.20 & $-$0&.30 \\
\hline
\end{tabular}\\[2pt]
\end{center}
\end{table}


\section{Observed spectra and determination of spectroscopic parameters}
\label{lab-results}

Prediction of the microwave spectra of HC$_5$N and HC$_7$N were very reliable based on 
earlier data \cite{HC5N_eQq_1978,HC5N_isos_v0_2004,HC7_9_11N_rot_eqQ_etc_2000,HC7N_mmW_2004}. 
Pickett's SPCAT and SPFIT programs \cite{spfit_1991} were used for prediction and fitting 
of the spectra, respectively. Each rotational level of HC$_5$N and HC$_7$N with $J > 0$ 
is split by spin coupling effects caused by the $^{14}$N nucleus ($I = 1$) into three 
HFS components. The rotational and spin angular momenta are coupled sequentially: 
\textbf{J} + \textbf{I}($^{14}$N) = \textbf{F}. At higher values of $J$, the strong HFS 
components are the ones conserving the spin orientation, i. e. those with 
$\Delta F = \Delta J$. The very narrow linewidths of a few kilohertz for an isolated 
line permits the corresponding HFS components to be resolved for all investigated 
transitions of HC$_5$N and HC$_7$N. However, the high sensitivity of the present HC$_5$N 
data allowed us to observe the weaker spin-reorienting HFS components with $\Delta F = 0$ 
for all transitions and the even weaker ones with $\Delta F = -\Delta J$ for the lowest 
two transitions ($J = 2 - 1$ and $3 - 2$). The transition frequencies obtained in the 
present investigation are summarized in Table~\ref{HC5N-exp-data} together with their 
quantum number assignments, uncertainties, and differences between experimental frequencies 
and those calculated from the global fit. The signal-to-noise ratios (S/N) of the spectral 
recordings of HC$_5$N were very high such that almost all lines were accurate to 0.2~kHz 
(1$\sigma$), see also Refs~\citenum{i-PrCN_rot_etc_2011,2-CAB_rot_2017}.


\begin{table}
\begin{center}
\caption{Rotational transitions of HC$_7$N with quantum numbers $J$ and $F$ recorded in the present study, 
         their frequencies (MHz), uncertainties Unc. (kHz), residuals O$-$C (kHz) between observed 
         transition frequencies and those calculated from the global fit.}
\label{HC7N-exp-data}
\begin{tabular}[t]{ccr@{}lr@{}lr@{}l}
\hline 
$J' - J''$ & $F' - F''$ &  \multicolumn{2}{c}{Frequency} &  \multicolumn{2}{c}{Unc.} &  \multicolumn{2}{c}{O$-$C} \\
\hline
 5 $-$ 4 &   4 $-$  3 &  5639&.9580 & 1&.0 & $-$0&.12 \\
         &   5 $-$  4 &  5640&.0104 & 1&.0 &    1&.19 \\
         &   6 $-$  5 &  5640&.0325 & 1&.0 &    0&.72 \\
 6 $-$ 5 &   5 $-$  4 &  6767&.9778 & 1&.0 &    0&.38 \\
         &   6 $-$  5 &  6768&.0107 & 1&.0 &    0&.72 \\
         &   7 $-$  6 &  6768&.0261 & 1&.0 & $-$0&.47 \\
12 $-$11 &  11 $-$ 10 & 13535&.9917 & 1&.0 & $-$0&.54 \\
         &  12 $-$ 11 & 13535&.9978 & 1&.0 & $-$1&.21 \\
         &  13 $-$ 12 & 13536&.0055 & 1&.0 &    1&.61 \\
\hline
\end{tabular}\\[2pt]
\end{center}
\end{table}


The very high quality of the HC$_5$N data prompted us to try to improve the data situation 
also for HC$_7$N. We recorded first data for the $J = 12 - 11$ and $6 - 5$ transitions 
which had been studied earlier \cite{HC7_9_11N_rot_eqQ_etc_2000}. The S/N were considerably 
lower than in the case of HC$_5$N, as was expected. Therefore, only the $\Delta F = \Delta J$
HFS components were detected with reasonable S/N even after long integration. In addition 
to remeasure transition frequencies of these two transitions, we recorded the $J = 5 - 4$ 
transition at 5640~MHz, which was the lowest in frequency available at that time. 
The HC$_7$N transition frequencies obtained in the present study are given with additional
information in Table~\ref{HC7N-exp-data}. The accuracies were 1~kHz (1$\sigma$) in 
all instances.

Our present HC$_5$N and HC$_7$N data were fit together with previous data. Since the number 
of spectroscopic parameters is small with respect to the total number of lines of each molecule, 
the quality of the reproduction of the lines of each source is a good indication for the 
accuracies of the lines. These were 10~kHz for the hyperfine free MW center frequencies 
from Alexander et al. \cite{HC5N_rot_dip_etc_1976} and 7~kHz for the mmW data from 
Winnewisser et al. \cite{HC5N_mmW_1982} and for the mmW and sub-mmW data from Bizzocchi 
et al. \cite{HC5N_isos_v0_2004} in the case of HC$_5$N. 
The accuracies were 1~kHz for the MW transition frequencies with HFS splitting from 
McCarthy et al. \cite{HC7_9_11N_rot_eqQ_etc_2000} and 10~kHz for those without HFS splitting 
in the MW region from Kirby et al. \cite{HC7N_rot_1980} and those in the mmW region from 
Bizzocchi et al. \cite{HC7N_mmW_2004} for the HC$_7$N molecule. The resulting spectroscopic 
parameters of HC$_5$N and HC$_7$N are given in Table~\ref{ground-state-parameter} together 
with the most recent previous data.

The line lists and spectroscopic parameters are available as supplementary material 
to this article; the line, parameter, and fit files along with additional files are 
deposited in the data section of the Cologne Database for Molecular 
Spectroscopy\footnote{Webaddress: https://cdms.astro.uni-koeln.de/classic/predictions/daten/} 
\cite{CDMS_2016}.


\begin{table*}
\begin{center}
\caption{Ground state spectroscopic parameters$^{a}$ (MHz) of HC$5$N and HC$_7$N in comparison 
         to the most recent previous data.}
\label{ground-state-parameter}
\renewcommand{\arraystretch}{1.10}
\begin{tabular}[t]{lr@{}lr@{}lcr@{}lr@{}l}
\hline 
 & \multicolumn{4}{c}{HC$_5$N} & & \multicolumn{4}{c}{HC$_7$N} \\
\cline{2-5} \cline{7-10} 
Parameter & \multicolumn{2}{c}{present} & \multicolumn{2}{c}{previous$^b$} & & \multicolumn{2}{c}{present} & \multicolumn{2}{c}{previous$^c$}\\
\hline
$B$                & 1331&.332691~(4) & 1331&.332687~(20) & &  564&.0011226~(75) &  564&.0011225~(44) \\
$D \times 10^6$    &   30&.1099~(8)   &   30&.1090~(15)   & &    4&.04117~(55)   &    4&.04108~(54)   \\
$H \times 10^{12}$ &    1&.642~(19)   &    1&.635~(29)    & &    0&.163~(10)     &    0&.163~(10)     \\
$eQq$(N)           & $-$4&.27680~(14) & $-$4&.242~(30)    & & $-$4&.275~(50)     & $-$4&.29~(16)      \\
$C \times 10^3$    &    0&.2896~(123) &     &             & &    0&.123          &     &              \\
\hline 
\end{tabular}\\[2pt]
\end{center}
$^a$ Numbers in parentheses are one standard deviation in units of the least significant figures. 
     The parameter $C$ of HC$_7$N was kept fixed to the value of a quantum chemical calculation.\\
$^b$ $B$, $D$, $H$ from Ref.~\cite{HC5N_isos_v0_2004}; $eQq$(N) from Ref.~\cite{HC5N_eQq_1978}.\\ 
$^c$ $B$, $D$, $H$ from Ref.~\cite{HC7N_mmW_2004}; $eQq$(N) from Ref.~\cite{HC7_9_11N_rot_eqQ_etc_2000}. 
\end{table*}


The large body of very accurate HC$_5$N transition frequencies from the present study not only 
led to an improvement in $eQq$(N) by more than two orders of magnitude but also resulted in a 
very well determined value for the nuclear magnetic spin-rotation coupling parameter $C$, 
the coefficient of \textbf{I$\cdot$J}. 
In addition, the uncertainties of $B$ and $D$ got reduced; the correlation of these parameters 
with $H$ is likely responsible for the improvement in the accuracy of $H$. The differences in 
the uncertainties of $eQq$(N) for HC$_7$N between present and previous data are modest; and we 
should point out that part of the improvement is caused by assigning an uncertainty of 1~kHz to 
the earlier FTMW data \cite{HC7_9_11N_rot_eqQ_etc_2000} whereas 2~kHz were reported initially. 
The nuclear spin-rotation parameter $C$ could not be determined for HC$_7$N, therefore, it was 
kept fixed to the value determined in a quantum chemical calculation. This treatment led to 
a change in $eQq$(N) from $-4.299 \pm 0.050$~MHz to $-4.275 \pm 0.050$~MHz.
The smaller uncertainty in the earlier $B$ value \cite{HC7N_mmW_2004} is caused by the fact that 
in that work unsplit (i.e. HFS free) transition frequencies were derived for the FTMW data 
\cite{HC7_9_11N_rot_eqQ_etc_2000} to which much smaller effective uncertainties were assigned. 
A more detailed discussion of the HFS parameters is presented in section~\ref{Discussion}.

\section{Application in astronomical observations}
\label{lsr-vel}

Emission lines of HC$_5$N may be prominent in dense cold molecular clouds such as TMC-1 
\cite{TMC-1_survey_2004}. Therefore, the very accurate rest frequencies from this and 
previous studies may be used to determine the local speed of rest in such a cloud with 
great accuracy, even more so as transitions occur with a spacing of $\sim$2660~MHz and 
because the $^{14}$N HFS splitting is usually resolved at low values of $J$. 
Table~\ref{lsr-velocity} lists some molecules which are often abundant in dense molecular 
clouds and have an at least moderately dense rotational spectrum and rest frequencies 
known very accurately at least in part. 
HC$_3$N in its ground vibrational and low-lying vibrational states \cite{HC3N_rot_IR_2017} 
or its isotopomers with one $^{13}$C \cite{HC3N-isos_rot_2001} or with 
D \cite{DC3N_rot_etc_2008} have larger rotational spacings, but may be used over a 
much wider frequency and rotational temperature ranges. The same applies to propyne 
\cite{MeC2_4H_rot_2008} and methyl cyanide 
\cite{MeCN_lamb-dip_2006,MeCN_v8le2_2015,MeCN-isos_rot_2009}. 
Carbonyl sulfide \cite{OCS_rot_2005,OCS_rot_2016} can be quite abundant in dense parts 
of molecular clouds. Sulfur dioxide is more suited for spectra of hot cores and hot corinos, 
the warm and dense parts of high- and low-mass star-forming regions, respectively. 
Its rotational spectrum \cite{SO2_OCS_rot_1984,SO2_MeOH_rot_1985,SO2_rot_1996,SO2_rot_2005} 
was also employed to calibrate rotational spectra of, e.g., H$_2$CS \cite{H2CS_rot_2008,H2CS_rot_2019}. 
Methanol is very abundant in many different types of molecular clouds, but most reported 
transitions frequencies have accuracies in the range of 50~kHz to 200~kHz; only limited 
data are available with better accuracies \cite{SO2_MeOH_rot_1985,MeOH_rot_2004,MeOH_HFS_2015}.


\begin{table*}
\begin{center}
\caption{Molecular species with a rich rotational spectrum suitable to determine the local speed 
         of rest with high precision, approximate upper frequency $\nu$ (GHz) for a certain accuracy, 
         approximate accuracy Acc. (kHz) and references Ref. and comments.}
\label{lsr-velocity}
\renewcommand{\arraystretch}{1.10}
\begin{tabular}[t]{lccl}
\hline 
Molecule & $\nu$ & Acc. & Ref. and comments\\
\hline
HC$_5$N   &   22 &   0.2   & This work.\\
HC$_5$N   &  500 &    7    & \cite{HC5N_isos_v0_2004} and references therein.\\
HC$_3$N   &  120 &   1.0   & \cite{HC3N_rot_IR_2017}.\\
HC$_3$N   &  900 &    10   & \cite{HC3N_rot_IR_2017} and references therein; including several excited states.\\
HC$_3$N   &  620 & 10$-$20 & \cite{HC3N-isos_rot_2001}; isotopic species with $^{13}$C.\\
CH$_3$CCH &  720 &   0.5   & \cite{MeC2_4H_rot_2008}; $K \le 12$.\\
CH$_3$CN  &  800 &   1.0   & \cite{MeCN_lamb-dip_2006}; $K \le 12$.\\
CH$_3$CN  & 1500 & 20$-$30 & \cite{MeCN_v8le2_2015}; $\varv _8 \le 2$, $K + l \le 12$.\\
CH$_3$CN  & 1200 & 20$-$50 & \cite{MeCN-isos_rot_2009}; isotopic species with $^{13}$C, $K \le 9$.\\
OCS       &  520 &   1.0   & \cite{OCS_rot_2005,OCS_rot_2016}; better than 0.1~kHz for several lines.\\
SO$_2$    &  120 &   2.0   & \cite{SO2_OCS_rot_1984,SO2_MeOH_rot_1985,SO2_rot_1996}; $\varv _2 \le 1$, also $\varv = 0$ of $^{34}$SO$_2$, many lines.\\
SO$_2$    & 2000 & 10$-$30 & \cite{SO2_rot_2005}; $\varv _2 \le 1$, several lines; many better than 50~kHz.\\
CH$_3$OH  &  120 &  1$-$10 & \cite{SO2_MeOH_rot_1985,MeOH_rot_2004,MeOH_HFS_2015} and references therein; selected lines, possibly up to 200~GHz.\\
\hline 
\end{tabular}\\[2pt]
\end{center}
\end{table*}

\section{Quantum chemical calculations}
\label{QCC}

Calculations for the equilibrium structure as well as the $^{14}$N nuclear quadrupole and 
spin-rotation coupling parameters were performed at the coupled-cluster (CC) level 
\cite{RevModPhys.79.291} using the coupled-cluster singles and doubles (CCSD) approach 
augmented by a perturbative treatment of triple excitations (CCSD(T)) 
\cite{CC+T_1989,doi:10.1063/1.460359,WATTS19921,doi:10.1063/1.471005} 
together with correlation consistent core-polarized valence (cc-pCVXZ, X = T, Q, 5, 6) 
\cite{cc-pVXZ_1989,core-corr_1995,WILSON1996339} basis sets. In the calculations of 
the spin-rotation tensors, perturbation-dependent basis functions, as described in 
Ref.~\cite{doi:10.1063/1.472143,doi:10.1080/002689797171346}, were used to ensure 
fast basis-set convergence.

Molecular equilibrium structures were obtained at the CCSD(T)/cc-pCVQZ level of theory, 
which was shown to yield molecular equilibrium structural parameters of very high quality 
for molecules carrying first-row atoms \cite{doi:10.1063/1.1357225}, which applies as well for 
the HCN \cite{doi:10.1063/1.1357225} and HC$_3$N \cite{DC3N_rot_etc_2008} systems studied here.
Following the extensive basis set study presented in Ref.~\cite{DC3N_rot_etc_2008} for the 
equilibrium structure and the nuclear quadrupole as well as spin-rotation coupling parameters 
of HC$_3$N (and DC$_3$N) we apply a similar procedure here. 
The $^{14}$N nuclear quadrupole coupling parameters of HCN, HC$_3$N, DC$_3$N, HC$_5$N, and HC$_7$N 
were determined at the CCSD(T)/cc-pCV5Z level augmented by relativistic corrections computed 
at the CCSD(T)/cc-pCVQZ level via second-order direct perturbation theory (DPT) 
\cite{doi:10.1063/1.2998300} and vibrational corrections evaluated using a perturbational approach 
(VPT2) \cite{doi:10.1021/j100552a013} as described in Ref.~\cite{doi:10.1063/1.1574314} for shieldings. 
Force fields at the MP2/cc-pVTZ level and nuclear quadrupole coupling parameters at the 
CCSD(T)/cc-pCVTZ level were employed for the vibrational corrections. The $^{14}$N spin-rotation 
parameters were calculated without further corrections at the CCSD(T)/cc-pCVQZ level. 
All correlated computations were carried out correlating all electrons. 
A value of 20.44(3)~mb was used for the $^{14}$N nuclear quadrupole moment \cite{TOKMAN199760}.

All quantum-chemical calculations were carried out using the CFOUR program package 
\cite{cfour}; the parallel version of CFOUR was used for some of the calculations 
\cite{harding_JChemTheoryComput_4_64_2008}. All results of calculations involving the coupled-cluster 
singles, doubles and triples (CCSDT) \cite{doi:10.1063/1.1589003,doi:10.1063/1.1462612} 
and coupled-cluster singles, doubles, triples and quadruples (CCSDTQ) 
\cite{doi:10.1063/1.1589003,doi:10.1063/1.1668632} approaches were obtained with 
the string-based many-body code MRCC \cite{mrcc} interfaced to CFOUR.

The resulting quantum chemically calculated $^{14}$N nuclear quadrupole coupling parameters of 
HCN to HC$_7$N are summarized in Table~\ref{comp_eQq_exp_ai_HC1-7N} together with experimental values.
The calculated and experimental $^{14}$N nuclear spin-rotation parameters are given in 
Table~\ref{comp_C_exp_ai_HC1-7N}.

\section{Discussion of hyperfine parameters}
\label{Discussion}

The nuclear quadrupole coupling parameters are usually interpreted in terms of bonding of the 
respective atom \cite{eQq_1955,eQq_2011}. It is not surprising that the $^{14}$N value of HCN is 
considerably different from that of HC$_3$N (and DC$_3$N), as shown in Table~\ref{comp_eQq_exp_ai_HC1-7N}. 
Unsurprisingly, the difference is small between HC$_3$N and HC$_5$N, and very close to zero between 
HC$_5$N and HC$_7$N. The calculated non-relativistic equilibrium values differ slightly from 
the experimental ground state values. The calculated vibrational corrections are modest, of order 
of one percent. Relativistic corrections are even smaller, as may be expected for a molecule 
containing only light atoms, but are not negligible at the level with which most of the 
experimental values have been determined. The agreement between calculated and experimental 
ground state values is excellent for HC$_3$N and DC$_3$N and very good for all others. 
The agreement in the case of HC$_7$N needs to be taken with some caution as the uncertainty 
of the experimental value is one order of magnitude larger than the deviation. 
Nevertheless, the value of $-$4275~kHz, determined with keeping the nuclear spin-rotation parameter 
fixed the value calculated by quantum chemical means, fits the trend better than the value 
$-$4299~kHz without this parameter in the fit, even though both values are compatible within 
the experimental uncertainties.

The agreement between calculated equilibrium nuclear spin-rotation parameters and the experimental 
ground state values in Table~\ref{comp_C_exp_ai_HC1-7N} is very good. The very small deviations for 
HCN and HC$_3$N may well be caused by the neglect of vibrational contributions; their determination 
could not be carried out with the programs at our disposal.


\begin{table*}
\begin{center}
\caption{Quantum chemically calculated non-relativistic equilibrium nuclear electric quadrupole parameters 
         $eQq$(N)$_e$ (kHz) of HCN, HC$_3$N, DC$_3$N, HC$5$N and HC$_7$N, relativistic (Rel.) and 
         vibrational (Vib.) corrections, and their sum (kHz) in comparison to experimental ground 
         state parameters and difference O$-$C (kHz) between experimental and calculated values.}
\label{comp_eQq_exp_ai_HC1-7N}
\renewcommand{\arraystretch}{1.10}
\begin{tabular}[t]{lr@{}lr@{}lr@{}lr@{}lr@{}lr@{}l}
\hline 
Molecule & \multicolumn{2}{c}{$eQq$(N)$_e$$^b$} & \multicolumn{2}{c}{Rel.$^c$} & \multicolumn{2}{c}{Vib.$^d$} 
& \multicolumn{2}{c}{Sum} & \multicolumn{2}{c}{Exptl.} & \multicolumn{2}{c}{O$-$C}\\
\hline
HCN     & $-$4668&.9 & 1&.8 & $-$32&.2 & $-$4699&.4 & $-$4707&.83~(6)$^e$  & $-$8&.4 \\
HC$_3$N & $-$4344&.5 & 3&.5 &    21&.0 & $-$4320&.0 & $-$4319&.24~(1)$^f$  &    0&.8 \\
DC$_3$N & $-$4344&.5 & 3&.5 &    22&.6 & $-$4318&.3 & $-$4318&.03~(30)$^g$ &    0&.3 \\
HC$_5$N & $-$4326&.0 & 3&.6 &    41&.7 & $-$4280&.7 & $-$4276&.80~(14)$^h$ &    3&.9 \\
HC$_7$N & $-$4326&.2 & 3&.6 &    52&.3 & $-$4270&.3 & $-$4275&.~(50)$^h$   & $-$4&.7 \\
\hline 
\end{tabular}\\[2pt]
\end{center}
$^a$ All quantum chemical results from this work; CCSD(T)/cc-pCVQZ structural parameters were used; 
     see also section~\ref{QCC} for further details. 
     Sources of experimental data are given separately. Numbers in parentheses of experimental data 
     are one standard deviation in units of the least significant figures.\\
$^b$ Calculated at the CCSD(T)/cc-pCV5Z level.\\
$^c$ Calculated at the CCSD(T)/cc-pCVQZ level using second-order direct perturbation theory.\\ 
$^d$ Calculated employing MP2/cc-pVTZ for the required force field and CCSD(T)/cc-pCVTZ for 
     the nuclear electric quadrupole parameters.\\
$^e$ Ref.~\cite{HCN_HFS_dip_1984}.\\
$^f$ Ref.~\cite{HC3N_HFS_dip_1985}.\\
$^g$ Ref.~\cite{DC3N_rot_etc_2008}.\\
$^h$ This work.
\end{table*}


While the differences between calculated and experimental nuclear quadrupole coupling 
parameters are relatively small, we note that this is most likely due to error compensation. 
The computed $^{14}$N nuclear electric quadrupole parameters are found to depend sensitively 
on the equilibrium structure used. Moreover, they show very slow convergence with respect 
to basis set size and correlation treatment, which will be discussed in the following.

Although structures obtained at the CCSD(T)/cc-pCVQZ level are quite close to equilibrium 
structures derived from experiment \cite{doi:10.1063/1.1357225}, a shift in the $^{14}$N 
value of HCN by $-$5.0~kHz is obtained at the CCSD(T)/cc-pCV5Z level upon going from the 
CCSD(T)/cc-pCVQZ equilibrium structure ($r_e({\rm CH}) = 106.554$~pm, and $r_e({\rm CN}) 
= 115.384$~pm) to the equilibrium structures derived from experiment ($r_e({\rm CH}) 
= 106.501(8)$~pm, and $r_e({\rm CN}) = 115.324(2)$~pm) \cite{doi:10.1063/1.463237}. 
It is assumed that similar deviations would be found if more accurate equilibrium structures 
would be available for HC$_3$N, HC$_5$N, and HC$_7$N. We would like to emphasize here that 
these corrections are rather small in relative terms, though not negligible in comparison 
with experimental accuracy, which is important to realize for the discussion that follows. 
Furthermore, the uncertainty of the value of the $^{14}$N quadrupole moment (20.44~mb) is 
$\pm$0.03~mb, which translates roughly to 0.15\% or values between $\pm$6.9 and $\pm$6.3~kHz 
for the molecules under study.

In the cases of HCN and HC$_3$N, it was possible to evaluate the equilibrium $^{14}$N values 
with an even larger basis. Separating the total $^{14}$N value into contributions from 
Hartree$-$Fock (HF) and CCSD(T) reveals on the one hand that the HF contribution still changes 
by $-$12.7~kHz and $-$12.0~kHz when increasing the basis set quality from cc-pCV5Z to cc-pCV6Z 
for HCN and HC$_3$N, respectively. On the other hand, the correlation contribution at this level 
changes from cc-pCV5Z to cc-pCV6Z by less than 1~kHz in the same direction. An equal shift 
of $-$12.0~kHz was found for HC$_5$N at the HF level by increasing the basis set level from 
cc-pCV5Z to cc-pCV6Z.

Considering higher-level correlation effects beyond CCSD(T) by employing fc-CCSDT/cc-pVTZ and 
fc-CCSDTQ/cc-pVDZ computations shows corrections that differ in sign but also in magnitude.
The difference between fc-CCSDT/cc-pVTZ and fc-CCSD(T)/cc-pVTZ is found to be $-$10.7 and $-$13.1~kHz 
for HCN and HC$_3$N, respectively. However, the corresponding differences between fc-CCSDTQ/cc-pVDZ 
and fc-CCSDT/cc-pVDZ are found to be about twice as large and positive with values of +20.5~kHz for HCN  
and +24.3~kHz for HC$_3$N.

Although the relativistic corrections obtained at the CCSD(T) level via second-order direct perturbation 
theory are relatively small, they are overestimated for all molecules but HCN by about 40\% if they are 
evaluated employing the smaller cc-pCVTZ basis set. In the case of HCN, they are overestimated by about 
70\% if the smaller basis is employed. However, for HCN the change when going from cc-pCVQZ to cc-pCV5Z
is only $-$0.2~kHz (about 10\%).

The vibrational correction obtained using VPT2 at the MP2/cc-pVTZ level for the required force field 
and at the CCSD(T)/cc-pCVTZ level for the nuclear quadrupole coupling parameters could be improved for HCN 
to CCSD(T)/cc-pCVQZ for the force field and CCSD(T)/cc-pCV5Z for the nuclear quadrupole coupling, 
which resulted in a modest change of $-$0.6~kHz. For HC$_3$N/DC$_3$N, improved VPT2 computations at the 
CCSD(T)/cc-pCVTZ level yielded vibrational corrections that are 5.3/5.1~kHz smaller than those 
reported in Table~\ref{comp_eQq_exp_ai_HC1-7N}.

Summarizing all the previously mentioned improvements for HCN, a total correction of $-$9.6~kHz 
is obtained, which would reduce the difference between observed and calculated from $-$8.4~kHz to +1.2~kHz. 
For HC$_3$N/DC$_3$N total corrections of $-$6.9/$-$6.7~kHz would be obtained, while the correction 
due to the uncertainty in the equilibrium structure remains unknown.

Since the discussed corrections can currently not be applied consistently to all 
molecules under study, it has to be assumed that similar compensation effects hold for the $^{14}$N nuclear 
quadrupole coupling parameters of HC$_5$N and HC$_7$N, which is supported by the {\it too good} agreement 
with experiment observed.

With respect to the $^{14}$N spin-rotation parameters, only little variations of the CCSD(T) results 
are found when increasing the basis set size. When going from cc-pCVQZ to cc-pCV5Z, changes of $-$0.028 
and 0.004~kHz are found for HCN and HC$_3$N, respectively. Employing again fc-CCSDT/cc-pVTZ and 
fc-CCSDTQ/cc-pVDZ computations to estimate the higher-level correlation effects in HCN, 
which has the largest spin-rotation parameter value, changes of +0.022 and $-$0.029~kHz are found 
for CCSDT and CCSDTQ, respectively, which almost cancel each other. 
However, comparing the resulting corrected values of 10.001~kHz (HCN) and 0.990~kHz (HC$_3$N) with
experiment, only an improvement for the value of HCN is observed, cf. Table~\ref{comp_C_exp_ai_HC1-7N}.


\begin{table}
\begin{center}
\caption{Experimental ground state nuclear magnetic spin-rotation parameters $C^{a}$ (kHz) 
         of HC$_5$N and HC$_7$N in comparison to experimental data for HCN, and HC$_3$N 
         and in comparison to equilibrium data from quantum chemical calculations (QCC).}
\label{comp_C_exp_ai_HC1-7N}
\renewcommand{\arraystretch}{1.10}
\begin{tabular}[t]{lr@{}lr@{}l}
\hline 
Molecule &  \multicolumn{2}{c}{Exptl.} & \multicolumn{2}{c}{QCC}\\
\hline
HCN$^b$     &  10&.13~(2)     & 9&.938 \\
HC$_3$N$^c$ &   0&.976~(3)    & 0&.986 \\
HC$_5$N$^d$ &   0&.2896~(123) & 0&.290 \\
HC$_7$N$^e$ &    &            & 0&.123 \\
\hline 
\end{tabular}\\[2pt]
\end{center}
$^a$ Numbers in parentheses of experimental data are one standard deviation in units of the least 
     significant figures. All quantum chemical results are from this work; calculations at the 
     CCSD(T)/cc-pCVQZ level for both structures and values of $C$; see also section~\ref{QCC}. 
     Sources of experimental data are given separately.\\
$^b$ Ref.~\cite{HCN_HFS_dip_1984}.\\
$^c$ Ref.~\cite{HC3N_HFS_dip_1985}.\\
$^d$ This work.\\
$^e$ The value was kept fixed in this work to the value of a quantum chemical calculation; 
     see also Table~\ref{ground-state-parameter}. 
\end{table}

\section{Conclusions}
\label{Conclusions}

Accurate transition frequencies of HC$_5$N and HC$_7$N were determined employing Fourier transform 
microwave spectroscopy. These data led to improvements of the spectroscopic parameters. In particular, 
we improved the accuracy of the $^{14}$N nuclear quadrupole coupling parameter of HC$_5$N considerably 
and that of HC$_7$N slightly. In addition, we determined for the first time an experimental value of the 
nuclear $^{14}$N nuclear spin-rotation parameter of HC$_5$N. Our quantum chemical calculations were 
able to reproduce the $^{14}$N hyperfine parameter of HCN to HC$_7$N very well. Vibrational 
corrections to nuclear quadrupole coupling parameters were crucial and relativistic corrections were 
important to reach nearly the accuracies with which these parameters were determined experimentally. 
Inclusion of a 
calculated, very small nuclear spin-rotation value of HC$_7$N had a remarkable effect on the value of the 
nuclear quadrupole coupling parameter. Finally, we pointed out that the HC$_5$N rest frequencies may be 
useful for determining the local speed of rest in dense molecular clouds if these lines are sufficiently 
strong. Other examples of molecular species with comparatively rich rotational spectra have been 
mentioned in addition.


\section*{Acknowledgments}

We are grateful to Peter F\"orster for initial measurements on HC$_5$N, to Holger Spahn 
for participation during the final experiments, and to Prof. Axel Klein and his group 
for preparing the HC$_3$N sample used in the present investigations. We thank the 
Laboratoire Europ{\'e}en Associ{\'e} de Spectroscopie Mol{\'e}culaire 'LEA-HiRes' for 
financial support. Additional funding was allocated by the Deutsche Forschungsgemeinschaft 
(DFG), in Cologne also within the Sonderforschungsbereich (SFB) 494. Further support was 
provided by the L\"ander Nordrhein-Westfalen, Niedersachen, and Hessen. The work in Mainz 
was supported by the DFG via grant GA 370/6-2 within the priority program SPP 1573. 
Our research benefited from NASA's Astrophysics Data System (ADS).

\appendix
\section*{Appendix A. Supplementary Material}

Supplementary data associated with this article can be found, in the online version, 
at ...



\bibliographystyle{elsarticle-num}
\bibliography{HC5N}






\end{document}